\documentclass[10pt, conference, compsocconf]{IEEEtran}
\usepackage{amssymb}
\usepackage[dvips]{graphicx}

\ifCLASSINFOpdf
 
\else

\fi

\begin{document}
\title{A Formal Semantic for UML 2.0 Activity Diagram based on Institution Theory}


\author{\IEEEauthorblockN{Amine Achouri}
\IEEEauthorblockA{Laboratory LaTICE\\
ESSTT\\
University of Tunis, TUNISIA\\
Amine.Achouri@fst.rnu.tn}
\and 
\IEEEauthorblockN{Leila Jemni ben Ayed}
\IEEEauthorblockA{Laboratory LaTICE\\
ESSTT\\
University of Tunis, TUNISIA\\
Leila.Jemni@fsegt.rnu.tn}
}


%Telephone: (800) 555--1212, Fax: (888) 555--1212}
\maketitle
\begin{abstract}
Giving a formal semantic to an UML Activity diagram (UML AD) is a hard task. 
The reason of this difficulty is the ambiguity and the absence of a precise formal 
semantic of such semi-formal formalism. 
A variety of semantics exist in the literature having tackled the aspects covered 
by this language. We can give as example denotational, functional and compositional semantics. 
To cope with the recent tendency which gave a heterogeneous semantic to UML diagrams, we aim to define an algebraic presentation 
of the semantic of UML AD. 
In this work, we define a formal semantic of UML 2.0 AD based on institution theory. 
For UML AD formalism, which is a graphical language, no precise formal semantic is given to it. 
We use the institution theory to define 
the intended semantic. Thus, the UML AD formalism will be defined in its own natural semantic. 

\end{abstract}

\begin{IEEEkeywords}
Institution theory; UML 2.0 Activity Diagram;  Formal semantic;

\end{IEEEkeywords}

% \fi
\IEEEpeerreviewmaketitle

\section{Introduction}
Model transformation is a critical process in software construction and development. 
As increasingly larger software systems are being developed, there is tendency to have solid and effective tools 
to automatize the software development. 
The specification of a software can be formal and (or) graphical. 
For graphical formalisms, we can mention as example UML models, UML class diagram, 
UML activity diagram and interaction diagram. 
For the formal ones, logic are increasingly used due to their mathematical background.
For example, Petri-net is used as a graphical and a formal specification formalism.
Logic is the language of formal methods such that theorem proving and model checking. 
To facilitate and to link graphical and formal language, there is a massive need to make generic 
techniques for the transformation of graphical models to formal notations. 
The use of logic is difficult for non familiar with logical
concepts and specification. As a result, there is a need to provide the possibility to make specifications in a modeling level. 

Stakeholders can begin with a graphical model (possibly with many system views).
Then, with an automatic and correct transformation they can produce a specification in a formal logic. 
In the context of logic, institution theory has emerged as a framework allowing their study and the 
different relation between them.

In our previous work \cite{Jemni10}, we used graph grammar to define an automatic transformation between UML AD and Event-B. 
Thanks to the notion of graph grammar, the automation aspect is given to the transformation. 
The semantic equivalence between source and target model is not proved. The reason is the absence of 
formal semantic for the source and the target formalism. 
To overcome this drawback, we use institution theory to make the required semantic 
for the source formalism which is UML AD. 

The first contribution aims to give institutional presentation of UML AD. 
In our knowledge, in the literature, no proven
institution for UML AD exists.
This institutional presentation define a formal semantic of UML AD.
In addition, this algebraic presentation of the source formalism will be a meta-level to study possible transformation to Event-B models \cite{Jemni10}. 
Thus, the study of some proprieties like model amalgamation and theory co-limits of this formalism will be enhanced \cite{Martins11}. 
Those notions play a key role in heterogeneous specification approaches. 
The UML AD institution may be used in a heterogeneous modeling language such UML diagram 
like in  \cite{Cengarle08c}. 

The paper is organized as follows: in section 2 we present the related works. 
Then in section 3, we recall institution definition. 
Section 4 shows how to prove that UML AD establish an institution. 
Section 5 give an example of UML AD model and make focus in its institution.
Finally, the last section concludes our work.

\section{Related works}
%\cite{Cengarle04} \cite{Cengarle08a} \cite{Cengarle08b} \cite{Lucanu05} \cite{Knapp10}

In literature, institution theory is largely used and studied. We have three category of works based on institution theory.

The first category is interested on the use of institution theory and its known concepts in the development of an heterogeneous
specification approaches. We mention 
the approach of the heterogeneous specification in the tool cafeOBJ \cite{Diaconescu_logicalfoundations}. 
This approach is based on a cube on eight logic and twelve projections (defined as a set of institution morphism and 
institution comorphism) \cite{Diaconescu_logicalfoundations}.
It's inspired by the semantic based on Diaconescu's notion of Grothendieck institution 
\cite{Diaconescu02grothendieckinstitutions}.
Another approach is developed in the work of Mossakowski \cite{Mossakowski05Habil} \cite{DBLP:conf/wadt/CodescuHKMRS10}. 
The heterogeneous logical environment developed by the author is formed by a number of logical systems. 
These logical systems are formalized as institutions linked with the concepts of institution morphism and comorphism. 

The second category of works focus on the use of institution theory in the specification of graphical formalism 
such as UML diagrams.
In this category, we mention the work present in \cite{Cengarle08c} \cite{Cengarle04} \cite{Cengarle08a} \cite{Cengarle08b}. 
The approach defined by Cengarle et al. aims to define a semantic for UML class diagram, UML interactions diagram
and OCL. Each diagram is described in its natural semantic because of the use of the algebraic formalization of each formalism.
In addition, relations between diagrams are expressed via institution morphism and comorphism. We note here that this approach
is inspired by Mossakowski works in the heterogeneous institution setting. 

The third category of works uses this theory for a specific intention and a precise case study. 
The work in \cite{Knapp10} is a good candidate in this category where authors defined a heterogeneous framework 
of services oriented system, using institution theory. 
Authors (in \cite{Knapp10}) aims to define a heterogeneous specification approach 
for service-oriented architecture (SOA). The developed framework consists of a several individual services specification  
written in a local logic. The specification of their interactions is written in a global logic. The two defined logics are described via
institution theory and an institution comorphism is used to link the two defined institution. This approach is inspired by the work 
of Mossakowski. Another work is developed in \cite{Lucanu05} where the authors propose to use institution to represent the logics 
underling OWL and Z. Then, they propose a formal semantic for the transformation  of OWL to Z specification via the use
of institution comorphism.  

Our proposed approach aims at first to give a semantic for UML AD via its representation as an institution. As a result, 
we propose to consolidate our approach given \cite{Jemni10}. 
Thus, with the defined semantic the transformation of UML AD model to an Event-B model can 
be semantically proven which means that the two model will be semantically equivalent.
It's clear that the approach we propose do not tackle the problematic of heterogeneous specification environment 
like \cite{Cengarle08c} and in \cite{Mossakowski05Habil}. The use of Event-B is argued with the following reasons:\\
\begin{itemize}
\item
Event-B is a formal method that supports interactive and automatic theorem proving. The resulted specification, after the transformation 
process, can be proved automatically. Event-B as a theorem prover is seeing a continuous improvement by industrial society.  
\item
With the notion of refinement, we can perform successive refinements to the Event-B model in order to obtain a pseudo code written
in declarative language.
\item
Thanks to the notion of composition supported in Event-B, we can define heterogeneous specification environment with different
graphical formalism. With the notion of composition, system described with heterogeneous specification can be composed 
and then proved formally. 
\end{itemize}

Our work is inspired form \cite{Mossakowski05Habil}. We are devoted to use 
UML AD as a formalism for applications modeling. This formalism will be represented as an institution. We intend to gain a formal semantic of UML AD thanks
to its algebraic categorical presentation. 

The version of UML AD used in this paper is 2.0. In literature, many approaches are proposed for the development of UML AD formal semantic. 
Recent works which treated the newest version are the work 
of St\"{o}rrle in \cite{Stoerrle2005} \cite{Stoerrle2004} \cite{Störrle05towardsa}. St\"{o}rrle provides a formal definitions
for the semantics of control-flow, procedure call, data-flow, and exceptions in UML
2.0 Activities. The defined semantic is inspired by Petri-net semantic. The choice of petri-net semantic by the authors is 
argued by the following reasons.
\begin{itemize}
 \item 
The standard claims that in the version 2.0 of UML AD \textit{Activities are redesigned to use a Petri-like semantics instead of state machines}.  
\item
Thanks to the formal foundation adequateness of Petri-net to give a formal semantic for UML AD
\item
In addition, in \cite{Stoerrle2004} St\"{o}rrle have shown how standard Petri-net tools may be applied to verify properties of 
UML 2.0 activity diagrams, using a Petri-net semantics.
\end{itemize}
In our paper, we will not use any intermediate semantic for UML AD such using Petri-net semantics. We provide a formal semantic of UML AD with mathematical notions
in term of categorical abstract presentation. 
We get profit from this categorical presentation the next benefit:
\begin{itemize}
\item
From this categorical presentation of the syntax and the semantic of UML AD, we can prove that UML AD can be written as an institution
\item
we can use the defined institution for an heterogeneous specification tools like \cite{Cengarle08c}
\item
Because we use Event-B as formal method for the verification of the UML AD 
we can use the concepts of institution comorphism and institution morphism to transform UML AD to Event-B 
\end{itemize}

% \section{The proposed approach} 
% 
% 
% \begin{figure}[!h]
% \begin{center}
% \includegraphics[width=11cm,height=7cm]{approach.eps}
% \end{center}
% \caption{Relations between source, target model and the model transformation in one side and the institution of UML AD and the institution of Event-B and the institution comorphism 
% in the other side}
% \label{fig:dessin13}
% \end{figure}
% 
% The approach we proposed is illustrated in fig \ref{fig:dessin13}:
% There are two models one is UML AD model and the other is Event-B model. The two models have a formal semantics in the institutions
%  UML AD institution and the event-B institution. The model transformation is applied between UML AD model and Event-B model. 
% This model transformation is backed by a semantic connection defined as an institution comorphism between the two defined institution
% (the institution of UML AD and the institution of Event-B institution). In the next section, we recall some important definition about 
% the theory institution that are the mathematical basic of our work.

\section{Logic as an Institution}

\noindent Institution is an abstract concept invented by Joseph Goguen and Rod Brustall because of the important variety of logics. 
It provide a basis for reasoning about software specifications independent of the choice of the
underlying logical system \cite{Goguen02}.

It offers an abstract theoretic presentation of logic in a mathematical way. 
An institution consists of notions of signatures, models, sentences, with a technical requirement, called the 'Satisfaction Condition', 
which can be paraphrased as the statement that 'truth is invariant under change of notation' \cite{Diaconescu08}. 
Modeling the signatures of a logical system as a category, we get the possibility to translate sentences and models
across signature morphisms.
The Satisfaction Condition is essential for reuse of specifications: 
it states that all properties that are true of a specification remain true in the context of another specification which imports 
that specification.

\begin{description}
 \item \textbf{Definition 1:}\\
\textit{An institution $I=(\mathbb{S}ig^{I}, \mathbb{S}en^{I}, \mathbb{M}od^{I}, \models^{I})$ consists of:}
\begin{itemize}
  \item 
  \textit{A category $\mathbb{S}ig^{I}$ whose objects are called signatures and the arrow are signature
   morphism.}
   \vspace{0.3cm}
  \item 
  \textit{A functor $\mathbb{S}en^{I}:\mathbb{S}ig^{I}\rightarrow \mathbb{S}et$, this functor map each signature $\Sigma$ to the set whose elements are called sentences constructed over that signature. Also Sen map each signature morphism to function between sentences.}
  \vspace{0.3cm}
  \item 
  \textit{A functor $\mathbb{M}od^{I}:(\mathbb{S}ig^{I})^{op}\rightarrow \mathbb{C}at$, this functor map each signature $\Sigma$ to the category of models of this signature. Also $\mathbb{M}od$ map each signature morphism to model homomorphism between models.}
  \vspace{0.3cm}
  \item \textit{A relation $\models_{\Sigma}^{I}$ giving for each sentences of a signature $\Sigma$ the models in which the sentences are true.}
\end{itemize}
\vspace{0.3cm}
 \end{description}
\textit{The relation $\models_{\Sigma}^{I}$ is called the satisfaction condition which can be interpreted like follows:\\
Given a signature morphism $\varphi$ :$\Sigma$ $\longrightarrow$ $\Sigma${'} in the institution I.}\\
\vspace{0.2cm} 
\textit{For each model $M{'}\in\mid \mathbb{M}od(\Sigma{'})\mid$ and e $\in \mathbb{S}en(\Sigma$)}\\ \vspace{0.2cm}
\textit{$\mathbb{M}od^{I}(\varphi)(M{'})\models_{\Sigma}^{I} e\Rightarrow M{'}  \models_{\Sigma{'}}^{I}   \mathbb{S}en^{I}(\varphi)(e)$ }
\vspace{0.2cm}
\\

\section{Using institution for the description of UML AD formalism} 

\subsection{Graphical Formalism}
\noindent UML activity diagrams (UML AD) are graphical notation developed by the OMG. 
It's used for the specification of workflow applications and to give details for an operation in software development. 
UML AD serve many purposes, during many phases of the software life cycle \cite{Stoerrle2004}. 
They are intended for being used for describing all process-like structures,
(business processes), software processes, use case behaviors, web services,
and algorithmic structures of programs.
UML AD are thus applicable throughout the whole software life cycle, which means  
during business modeling, acquisition, analysis, design, testing, and
operation, and in fact in many other activities.
◦Thus, they are intended for usage not just by Software-Architects and
Software-Engineers, but also by domain specialists, programmers, administrators and so on.
Some works in the literature use to define an institution for UML diagrams, we mention \cite{Cengarle08c} \cite{Cengarle04} \cite{Cengarle08a} 
\cite{Cengarle08b}. The cited works is devoted to define three
institution for respectively UML Class diagram, UML Interactions Diagram and OCL. In our paper, the semantic of UML AD
will be based on the works of H. St\"{o}rrle. As we say in the previous section, the considered work is the more recent 
and relevant work in this context conformed with the standard.
 
\begin{figure}[!h]
\begin{center}
\includegraphics[width=8.5cm,height=8.5cm]{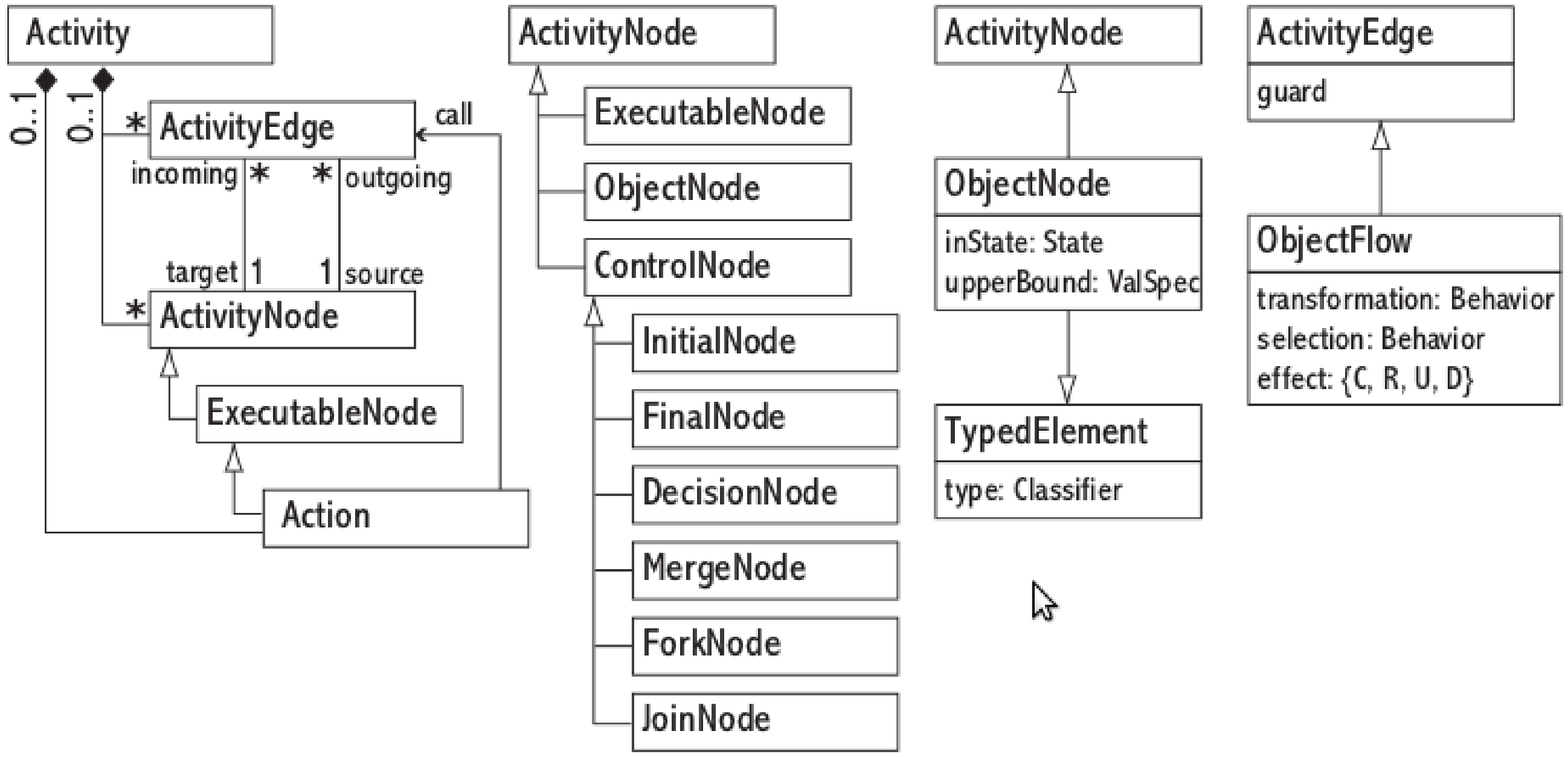}
\end{center}
\caption{A portion of the meta model of UML AD (as it is defined in the standard).}
\label{fig:dessin11}
\end{figure}

With the version 2.0 of UML AD, the meta-model for Activities has been redesigned from
scratch (fig \ref{fig:dessin11}). The main concept underlying Activity Diagrams is now called Activity \cite{Störrle05towardsa}.  
The meta-model defines six levels increasing expressiveness. 
The first level (“Basic Activities”) already includes control flow and procedurally
calling of subordinate Activities by Activity Nodes that are
in fact Actions (see fig \ref{fig:dessin11}). This paper is restricted to Basic Activities.
Readers may refer to \cite{Stoerrle2004} \cite{Stoerrle2005} \cite{Störrle05towardsa} 
for more details about the syntax and the semantic of UML AD.

Next, we will prove that UML AD formalism can be written as an institution. 

\subsection{The syntax of UML AD}

Activity as defined in \cite{Stoerrle2005} is the coordination of elementary actions or it consists of one atomic action.
Besides, given a class diagram, methods are functions that uses attributes of the considered class. Then, class diagram methods are
functions or operations that changes the state of an object (defined as an instance of the considered class). 
In this two cases, we consider an activity as a method of a class in UML class diagram or we consider 
an activity as a coordination of one action or more. As result, we can define a relation of hierarchy. 
This relation is defined between two activities Activity A and Activity B.\\
An activity hierarchy $\mathbb{A}$ written as $\mathbb{A} = (\textit{A}, \preccurlyeq_{\textit{A}} )$ 
is a partial order with a set of activity names $\textit{A}$ and a subclass relation 
$\preccurlyeq_{\textit{A}} \subseteq \textit{A}\times \textit{A}$. \\
Given an activity hierarchy  $\mathbb{A} = (\textit{A}, \preccurlyeq_{\textit{A}} )$,
a \textit{A}-activity domain is a \textit{A}-indexed family
$\mathbb{N}= (\textit{N}_{\textit{a}} )_{\textit{a}\in \textit{A}}$ of sets of activity with 
$\textit{N}_{\textit{a}} \subseteq \textit{N}_{\textit{a{'}}}$ if $\textit{a} \preccurlyeq_{\textit{A}} \textit{a{'}}$. 
We aim to prove that the Activity hierarchies can be formalized as a category which can be done 
via it's formalization as a Grothendieck construction and also as a monad.
The two presentations of Activity hierarchies as Grothendieck construction and as a monad are shown in \cite{Cengarle08b}
(with replacing class hierarchies with Activity hierarchies).

An UML AD signature consists of a pair $\Sigma=(\mathbb{A}, \textit{E})$ where $\mathbb{A}$ is the activity hierarchy
and $\textit{E}$ is the set of Activity Edges.\\
Given a signature $\Sigma=(\mathbb{A}, \textit{E})$ with $\mathbb{A} = (\textit{A}, \preccurlyeq_{\textit{A}} )$, we define 
a set \textit{\textbf{T}} of atomic formulas over $\Sigma$ by:\\
\vspace{0.2cm}
\textbf{\textit{\textbf{T}} :=skip $\mid$ seq(C,e,D)}% $\wedge$ \textit{\textbf{T}}}\\
with e $\in \textit{E}$ and C, D $\in$ \textit{A},  \\

Given UML AD signatures $\Sigma_{1}=(\mathbb{A}_{1},$ \textit{$E_{1}$}) and $\Sigma_{2}=(\mathbb{A}_{2},$ \textit{$E_{2}$}).

We define a UML AD signature morphism $\varphi: \Sigma_{1}\longrightarrow \Sigma_{2}$ as a morphism that maps Activity node names to Activity node names and maps
Activity Edges to Activity Edges. We note here that Activity node can be one of the following node:
\begin{itemize}
 \item 
EN: The set of Executable Nodes (i.e. elementary Actions);
\item 
IN or FN : The Initial Nodes or the Final Nodes 
\item
BN: the set of branch nodes, including both Merge Nodes and Decision Nodes 
\item
CN: the set of concurrency nodes, subsuming Fork Nodes and Join Nodes;
\item
ON: the set of Object Nodes;
\end{itemize}
As for Activity Edges may be a pair AE, OF , where:
\begin{itemize}
 \item 
AE: the set of plain Activity Edges between Executable Nodes and Control Nodes;
 \item 
OF: the set of Object Flows between Executable Nodes and Control Nodes on the one hand, and Object Nodes on the other.
\end{itemize}
Signature morphism extend to atomic formulas over $\Sigma_{1}$ as follows:\\
\vspace{0.2cm}
\textbf{$\varphi(skip)=skip$}\\
\textbf{$\varphi$(seq($C_{1}$,$e_{1}$,$D_{1}) $%\wedge$ \textit{\textbf{$T_{1}$}})= 
=seq($\varphi(C_{1})$,$\varphi(e_{1})$,$\varphi(D_{1}))$}
%\wedge$ \textit{\textbf{$\varphi(T_{1})$}}}

Let $\Sigma=(\mathbb{A},$ \textit{E}) be an UML AD signature. X =($X^{a})_{a\in A}$. 
The language of propositional ($\Sigma$,X) formulas has the below form:\\
\textbf{\textit{\textbf{T}} :=skip $\mid$ seq(C,e,D)}.\\
The language of first order ($\Sigma$,X) formulas has the form:\\
\textbf{$\phi$::=\textit{T} $\mid$ \textit{\textbf{T}}=\textit{\textbf{T}} $\mid \neg \phi \mid \phi \wedge \phi
\mid \phi \vee \phi \mid \phi \Longrightarrow \phi \mid \phi \Leftrightarrow \phi \mid (\exists x) \phi \mid (\forall x) \phi $}.\\
$\Sigma$ sentences are closed formulas defined on ($\Sigma$,X) formulas.

\subsection{The semantic of UML AD}

In the standard, the semantic of UML AD is determined by a path expressing the trace of the execution.
For the execution, a token will move from the Initial Activity Node To the Final Activity Node \cite{Stoerrle2004}.
Each Activity has its role in AD execution \cite{Störrle05towardsa}.
First of all, a token in the Initial Node means the beginning of the execution of UML AD. 
Then, the trace of the token will be defined by the outgoing edges of the Initial node. 
When a token arrive to an Executable Node, it will trigger the Action or the operation in this node. 
For the Join Node, “if there is a token offered on all incoming edges, then a token are offered on the outgoing
edge”. A Fork Node means that, “when an offered token is accepted on all the outgoing edges,
duplicates of the token are made and one copy traverses each edge”.
 In the case of Merge Node and Decision Node, every edge (s) respectively incoming and outgoing  
is associated to a condition determining the condition 
of the activation of this edge. 
For Merge Node, “if there is a token offered to only one of the incoming edges where the condition is true (it's a sufficient condition), 
then a token are offered on the outgoing edge of the Merge Node”. 
A Decision Node means that in the outgoing edge where the condition is true, an offered token will traverses this edge. 
A token that traverses a Object Node means the availability of the object (variable) needed to the execution of the coming activity.

Given a UML signature $\Sigma=(\mathbb{A},$ \textit{E}) with $\mathbb{A} = (\textit{A}, \preccurlyeq_{\textit{A}} )$,
a structure \textit{I} for $\Sigma$ is a triple \textit{I}=($\mathbb{N},\mathbb{E},\mu$) where 
$\mathbb{N}$=($\textit{N}^{a})_{\textit{a}\in \textit{A}}$ is an Activity domain for $\mathbb{A}, \mathbb{E}$ a domain 
of edges and $\mu$ : \textit{E} $\longrightarrow \mathbb{E}$ is an interpretation function for edges.
Given a variable C a valuation $\beta$ for C in \textit{I} assigns values to variables. This means:\\
$\beta: C\longrightarrow \textit{N}^{a}$

A sub-signature $\Sigma{'}=(\mathbb{A}{'}, \textit{E}{'}) \subseteq \Sigma $ with 
$\mathbb{A}{'}= (\textit{A}{'}, \preccurlyeq_{\textit{A}{'}} )$ induces a set of traces 
\textit{T}($\Sigma{'}$,\textit{I}) defined as follows:\\
\vspace{0.3cm}
\textit{T}($\Sigma{'}$,\textit{I})=\{$e_{1}.e_{2}..e_{n}\mid i\in\{1,...,n\}, 
e_{i}=seq(C_{i},e_{i},D_{i}), C_{i}, D_{i}\in \mathbb{A}{'} and e_{i}\in  \textit{E}{'} $\}\\
The set of $\mathbb{T}$(\textit{I})of all traces is defined as :\\
\vspace{0.3cm}
$\mathbb{T}$(\textit{I})=\{$e_{1}.e_{2}..e_{n}\mid i\in\{1,...,n\}, 
e_{i}=seq(C_{i},e_{i},D_{i}) and C_{i}, D_{i}, e_{i}\in  \textit{I} $\}\\

The set $\Theta(\textit{T},\beta)$ of traces of an atomic formula \textit{T} over $\Sigma$ in the structure \textit{I} under the 
valuation $\beta$ are inductively defined as follows:\\
T:=skip $\Longrightarrow \Theta(\textit{T},\beta)$=\{$\varepsilon$\}\\
T:=seq(C,e,D) $\Longrightarrow \Theta(\textit{T},\beta)=\{seq(\beta(C),\mu(e),\beta(D))\}$

\textbf{\textit{\textbf{T}} :=skip $\mid$ seq(C,e,D)} %$\wedge$ \textit{\textbf{T}}}\\
with e $\in \textit{E}$ and C, D $\in$ \textit{A},  \\

\subsection{The satisfaction condition under the UML AD institution}

Let $\Sigma_{1}=(\mathbb{A}_{1}, \textit{E}_{1})$ and $\Sigma_{2}=(\mathbb{A}_{2}, \textit{E}_{2})$ be two UML AD signatures,  
an UML AD signature morphism $\varphi: \Sigma_{1}\longrightarrow \Sigma_{2}$, two structure $\textit{I}_{1}$ a $\Sigma_{1}$-structure 
and $\textit{I}_{2}$ a $\Sigma_{2}$-structure defined as $\textit{I}_{1}=(\mathbb{N}_{1},\mathbb{E}_{1},\mu_{1}$) and
$\textit{I}_{2}=(\mathbb{N}_{2},\mathbb{E}_{2},\mu_{2}$).
Semantic invariance under the change of notation is formulated as 
$\Theta_{\textit{I}_{2}}(\varphi(\textit{T}_{1}),\beta_{2})=\Theta_{\textit{I}_{1}}(\textit{T}_{1},\beta_{1})$ for any atomic
formula $\textit{T}_{1}$ over $\Sigma_{1}$.
This can be shown by induction on the structure of $\textit{T}_{1}$.\\
$\Theta_{\textit{I}_{2}}(\varphi(skip),\beta_{2})=\{\varepsilon\}=\Theta_{\textit{I}_{1}}(\varphi(skip),\beta_{1})$\\
$\Theta_{\textit{I}_{2}}(\varphi(seq(C,e,D) ),\beta_{2})=
\Theta_{\textit{I}_{2}}(seq(\varphi(C),\mu(e),\varphi(D))(skip),\beta_{2})=\\
\{seq(\beta_{2}(\varphi(C)),\beta_{2}(\mu(e)),\beta_{2}(\varphi(D))\}=
\{seq(\beta_{1}(C),\beta_{1}(e),\beta_{1}(D)\}=
\Theta_{\textit{I}_{1}}(\textit{T}_{1},\beta_{1})$

Also we have \textit{T}($\varphi(\Sigma_{1}),\textit{I}_{2})=\textit{T}(\Sigma_{1},\textit{I}_{1})$

\subsection{The institution of UML AD}

After this theoretic study of UML AD, we can prove that it form an institution. We can immediately observe 
that institutional presentation rely heavily on the institution of First Order Logic. 
 
\begin{description}
\item \textbf{Proposition 1:}\\
\textit{UML Activity Diagram form an Institution presented as below:}
\begin{itemize}
\item
\textit{Signatures declares Activity Nodes names, Edges Nodes names.}
\vspace{0.2cm}
\item
\textit{Sentences are closed formulas where well formed formulas combines atomic formulas using the conjunction, 
negation, universal quantification and equality of variables. The atomic formulas associated to UML AD are UML AD branch 
(connection between Activity Node names) and it's composition using the operator seq.}
\vspace{0.2cm}
\item
\textit{Model interprets each signature as follows:}
\vspace{0.2cm}
\begin{itemize}
\item \textit{Each activity node (depending to Activity Node type) as:  } 
\begin{itemize}
 \item 
\textit{An instance of Executable Nodes  if it denote the set EN.}
\item 
\textit{A truth valuation if it is Initial Nodes or the Final Nodes.}  
\item
\textit{A valuation to true or false depending to the condition on the branch nodes (including both Merge Nodes and Decision Nodes).}
\item
\textit{A valuation to true when it denote a concurrency nodes, subsuming Fork Nodes and Join Nodes.}
\item
\textit{An instance of object or an attributes on a Object for Object Nodes.}
\end{itemize}
\item \textit{As for Activity Edges the interpretation: }
\begin{itemize}
 \item 
\textit{An instance showing the end of execution of the Activity Node (where this edge is defined as the outgoing connection) 
and the beginning of the execution of another Activity (where this edge is defined as the incoming connection).}
\end{itemize}
\end{itemize}
\end{itemize}  
\end{description}

\section{Exemple of UML AD model}

\begin{figure}[!h]
\begin{center}
\includegraphics[width=10cm,height=8.5cm]{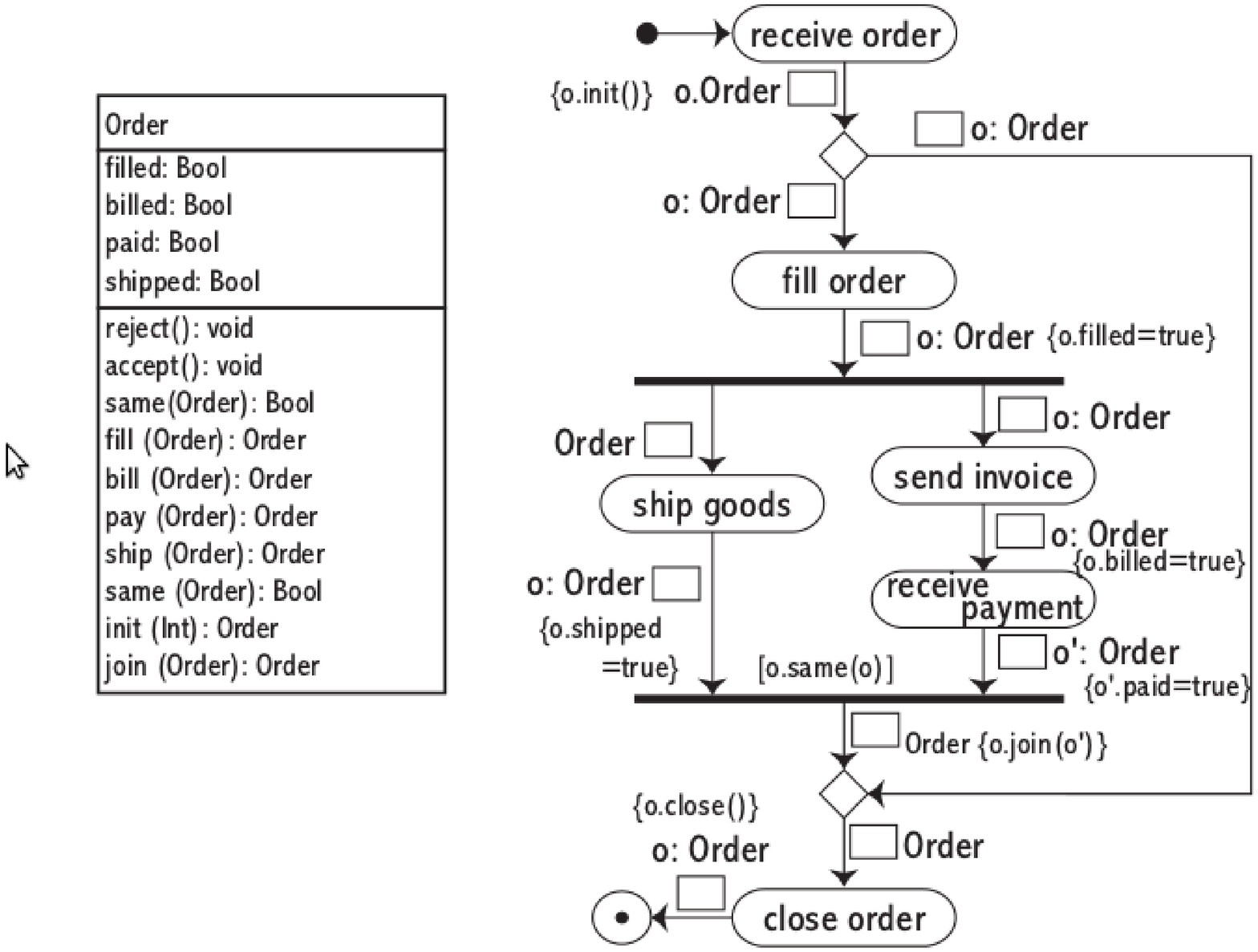}
\end{center}
\caption{An example of UML AD model(\cite{Stoerrle2005})}
\label{fig:dessin12}
\end{figure}

The example of the figure \ref{fig:dessin12} is presented in (\cite{Stoerrle2005}).
It represent an UML AD model and UML class diagram. The later contain the different action(method) used in the UML AD model.
From the categorical theoretic presentation of UML AD in the previous subsection, we can identify the signatures, the sentences and 
the interpretation of the example \ref{fig:dessin12}.

For the example (fig \ref{fig:dessin12}) the signatures declares Activity Node names Initial Node, receive order, fill order, ship goods, 
send invoice
receive payment, close payment, Final node, And Split, Or Split, And Join and Or Join. And split denote a subsuming Fork. 
Or Split denote a Decision Node. And Join denote a Join Nodes. Or Join denote a Decision Node.
As for edges, the example declares e1, e2 e3, e4, e5, e6, e7, e8, e9, e10, e11, e12, e13.
The sentence presented by the above example is the following closed formulas:\\
\textbf{ seq(Initial Node,e1,receive order) $\wedge$ seq(receive order,e2,Or Split) $\wedge$ \\seq(Or Split,e3,Or Join) $\wedge$ 
seq(Or Split,e4,fill order) $\wedge$\\ seq(fill order,e5,And Split) $\wedge$ seq(And Split,e6,ship goods) $\wedge$\\ 
seq(And Split,e7,send invoice) $\wedge$ seq(ship goods,e8,And Join) $\wedge$\\ seq(send invoice,e9,receive payment)
 $\wedge$ seq(receive payment,e10,And Join)$\wedge$\\ seq(And Join,e11,Or Join)$\wedge$ seq(Or Join,e12,close payment) $\wedge$\\
seq(close payment,e13,Final node)}.\\
%An interpretation of UML AD maps 

\section{Conclusions}

In our paper, we investigated the use of institution theory in a modeling formalism. 
We are motivated by the fact that we want to borrow the verification of system requirement and UML AD properties to Event-B.
In other terms, we aim to verify properties inexpressible in UML AD model with the theorem prover Event-B. The institution
of UML AD work as a meta-modelling language for this formalism. In addition, UML AD model conformance with the meta-model (formalism)
will be seen as a verification of the syntax correctness in the framework of UML AD institution. 
The defined syntax for UML AD don't address the whole syntax such it's defined in the standard. 
As future work, we aim to add more aspects for the UML AD institution. Then, We intend to prove an institution of Event-B and 
an institution comorphism from UML AD institution to Event-B institution. 
Thus, the semantic equivalence between source and target model will full preserved.

\end{document}